\documentclass[12pt,a4paper]{article}
\usepackage[dvips]{graphicx}
\begin{document}
\title{About agreement of PYTHIA and the experimental results in $e^+e^-$ annihilation to hadrons}
\author{N.\,V. Radchenko\thanks{nvrad@mail.ru}\\
        Novgorod State University, B. S.-Peterburgskaya Street 41,\\
        Novgorod the Great, Russia, 173003}
\date{}
\maketitle

\begin{abstract}
The experimental charged particles multiplicity distributions in
$e^+e^-$ annihilation to hadrons are compared with the
distributions obtained by PYTHIA. The ratio $\chi^2$/degrees of
freedom is calculated for 6 energies at $\sqrt{s}$ 14 -- 206.2
GeV. The necessity of more subtle tuning of PYTHIA at the energy
of $Z^0$ peak is discussed.

\end{abstract}
\section{Introduction}
In the last decades both experimentalists and theorists have
applied significant forces to understand character of hadron
interactions at high energies. As result many aspects of hadron
processes at high energies  became clear not only qualitatively
but also quantitatively. The last corresponds to quantum
chromodynamics which was able to describe quantitatively many the
simplest processes occurring at short distances. However, it is
impossible to count on QCD calculations of the multiple hadron
production, these processes have enormous number of degrees of
freedom. Thus one has to use different model approaches. A
quantitative description of high energy processes by Monte Carlo
generators can be considered as one of these approaches. Monte
Carlo generators are perspective method of theoretic study at all
energies of the existing accelerators including LHC, since any
other approaches, theoretical as the BFKL theory or
phenomenological as the Regge theory, are only valid to the values
of $1/\ln\sqrt{s}$ order of magnitude, where $s$ is the square of
total center-of-mass energy. Since even for LHC $\ln\sqrt{s}\simeq
19$, then it is necessarily to take into account the next to
$1/\ln\sqrt{s}$ corrections what is an inexecutable task at the
most cases. Therefore it is essentially to construct such MC
generator which will describe well processes both in hard and soft
regions.

Amongst generators describing well hard processes there is known
PYTHIA~\cite{bib:1}. Still soft processes, i.e. processes with
small transverse momenta, are included in PYTHIA only on basis of
the simplest phenomenological models, the values of cross sections
are not generated. The multiparticle processes are described with
the simplest pomeron model in which only one string is stretched
between quark and diquark. Pomeron cuts are not taken into
account. The weakness of PYTHIA is also in absence of the AGK sum
rules~\cite{bib:2}. An oscillation structure in the multiplicity
distribution function $P_n=\sigma_n/\sigma_{tot}$, where
$\sigma_n$ is the topological cross section of $n$ hadrons
production, $\sigma_{tot}$ is the total cross section of
hadron-hadron scattering,  that can be observed at the LHC
energies, follows from the AGK rules. Since PYTHIA is the most
advanced programm then the future MC generator describing both
hard and soft regions, in our opinion, should be based on PYTHIA.

In the present work we will study how multiplicity distributions
obtained with PYTHIA agree with experimental data.

\section{The charged particles multiplicity distribution in $e^+e^-$
annihilation to hadrons and PYTHIA predictions} We consider the
process of $e^+e^-$ annihailation at 6 values of total
center-of-mass energy from 14 to 206.2 GeV. Wide coverage of the
experimental data and  sufficient statistical provision of them
were criteria for the choice.

Electron-positron annihilation to hadrons was simulated by PYTHIA
at $\sqrt{s}$ 14, 29, 34.8, 91.2, 188.6, 206.2 GeV. The decay
process of $\gamma^\ast\left(Z^0\right)$ was considered without
the initial state radiation, values of the other parameters were
set by default. There were generated one million events for every
energy, and the charged particles multiplicity distributions were
analyzed.

\newpage
\subsection{$\sqrt{s}=14$ GeV}
Experimentally there were obtained 2\,704 events
(TASSO)~\cite{bib:3}. The experimental values of $P_n$ and the
corresponding probabilities obtained by PYTHIA are shown in Fig.1.
Both statistical and systematical errors are  included. The
$\chi^2$ is presented in Table 1.

\begin{center}
\includegraphics[height=80mm,width=120mm]{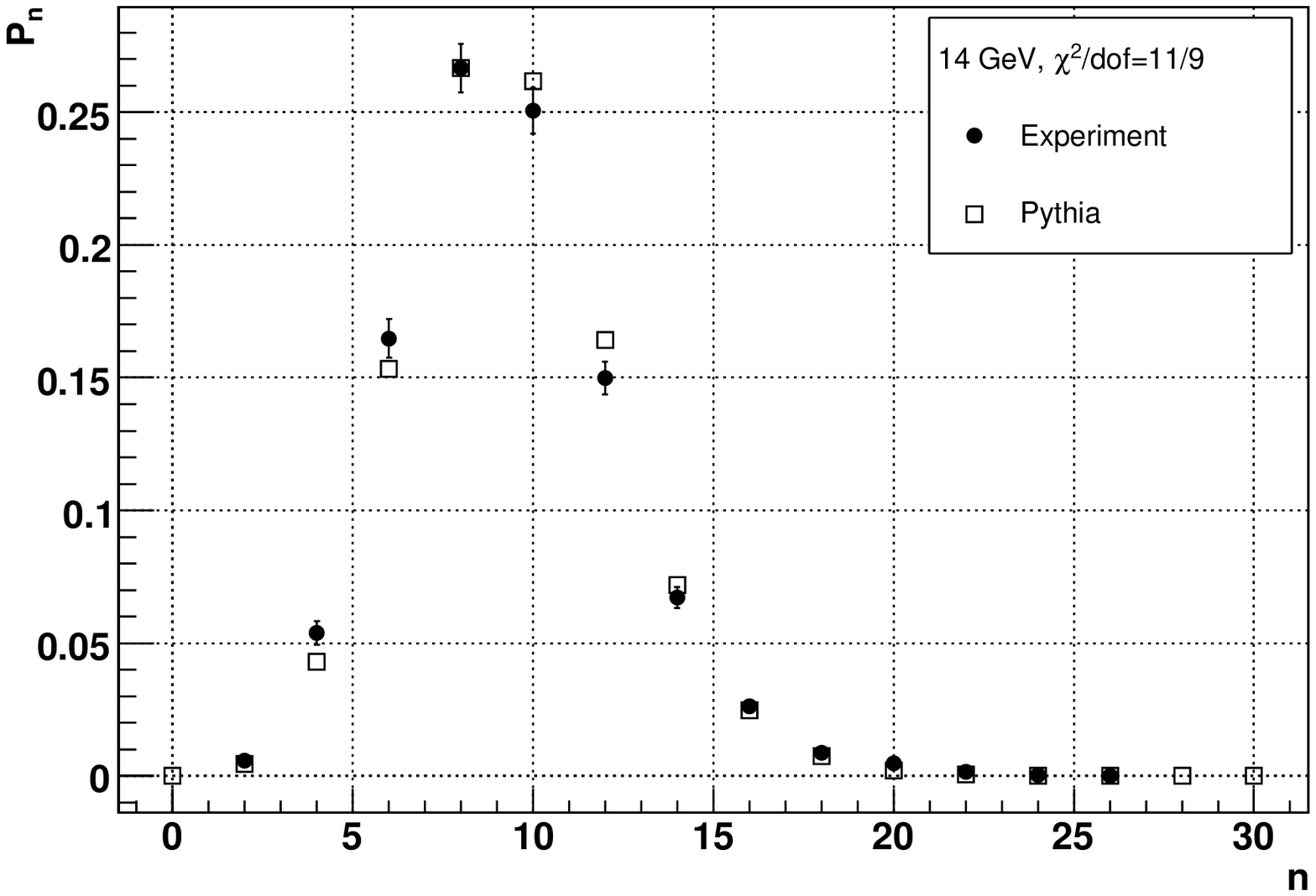}
\end{center}
\begin{center}
\textbf{Fig.1.}
\end{center}

\begin{center}
\begin{tabular}{|c|c|c|c|}
\multicolumn{4}{l}{\textbf{Table 1.} $\sqrt{s}=14$ GeV}\\
\hline$n$&number of events, experiment&number of events, PYTHIA&$\chi^2$ in bin\\
\hline0 -- 2&15.64&12.20&0.33\\
\hline4&145.98&116.08&3.43\\
\hline6&445.47&414.55&1.20\\
\hline8&720.66&720.78&0.00\\
\hline10&677.63&707.34&0.71\\
\hline12&405.05&443.66&2.06\\
\hline14&181.75&194.74&0.55\\
\hline16&70.79&66.74&0.15\\
\hline18&23.53&20.51&0.28\\
\hline20 -- 30&17.51&7.40&2.64\\
\hline \multicolumn{4}{|c|}{total $\chi^2$/degrees of freedom = 11/9}\\
\hline
\end{tabular}
\end{center}

\newpage
\subsection{$\sqrt{s}=29$ GeV}
Experimentally there were obtained 29\,649 events
(HRS)~\cite{bib:4}. The experimental values of $P_n$ and the
corresponding probabilities obtained by PYTHIA are shown in Fig.2.
Both statistical and systematical errors are included. The
$\chi^2$ is presented in Table 2.

\begin{center}
\includegraphics[height=80mm,width=120mm]{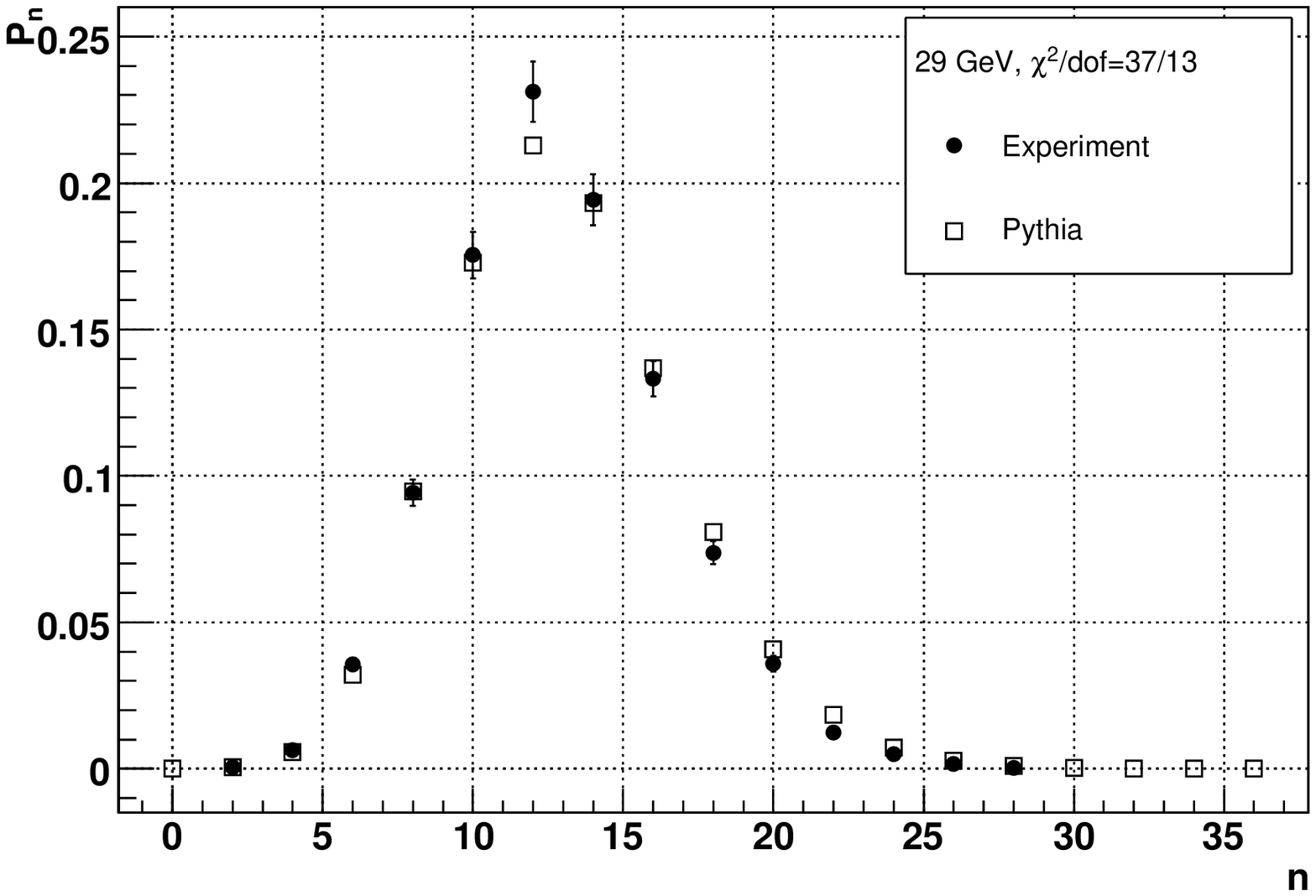}
\begin{center}
\textbf{Fig.2.}
\end{center}
\end{center}

\begin{center}
\begin{tabular}{|c|c|c|c|}
\multicolumn{4}{l}{\textbf{Table 2.} $\sqrt{s}=29$ GeV}\\
\hline$n$&number of events, experiment&number of events, PYTHIA&$\chi^2$ in bin\\
\hline0 -- 2&14.82&12.87&0.05\\
\hline4&186.79&169.59&0.08\\
\hline6&1055.50&952.57&1.98\\
\hline8&2795.90&2811.24&0.01\\
\hline10&5200.43&5122.86&0.10\\
\hline12&6854.85&6313.39&2.91\\
\hline14&5760.80&5724.38&0.02\\
\hline16&3949.25&4057.80&0.32\\
\hline18&2188.10&2397.37&2.73\\
\hline20&1064.40&1212.53&2.82\\
\hline22&367.65&545.37&11.70\\
\hline24&148.25&213.92&4.70\\
\hline26&50.40&79.13&2.76\\
\hline28 -- 36&11.86&35.99&7.29\\
\hline \multicolumn{4}{|c|}{total $\chi^2$/degrees of freedom = 37/13}\\
\hline
\end{tabular}
\end{center}

\newpage
\subsection{$\sqrt{s}=34.8$ GeV}
Experimentally there were obtained 52\,832 events
(TASSO)~\cite{bib:3}. The experimental values of $P_n$ and the
corresponding probabilities obtained by PYTHIA are shown in Fig.3.
Both statistical and systematical errors are included. The
$\chi^2$ is presented in Table 3.

\begin{center}
\includegraphics[height=80mm,width=120mm]{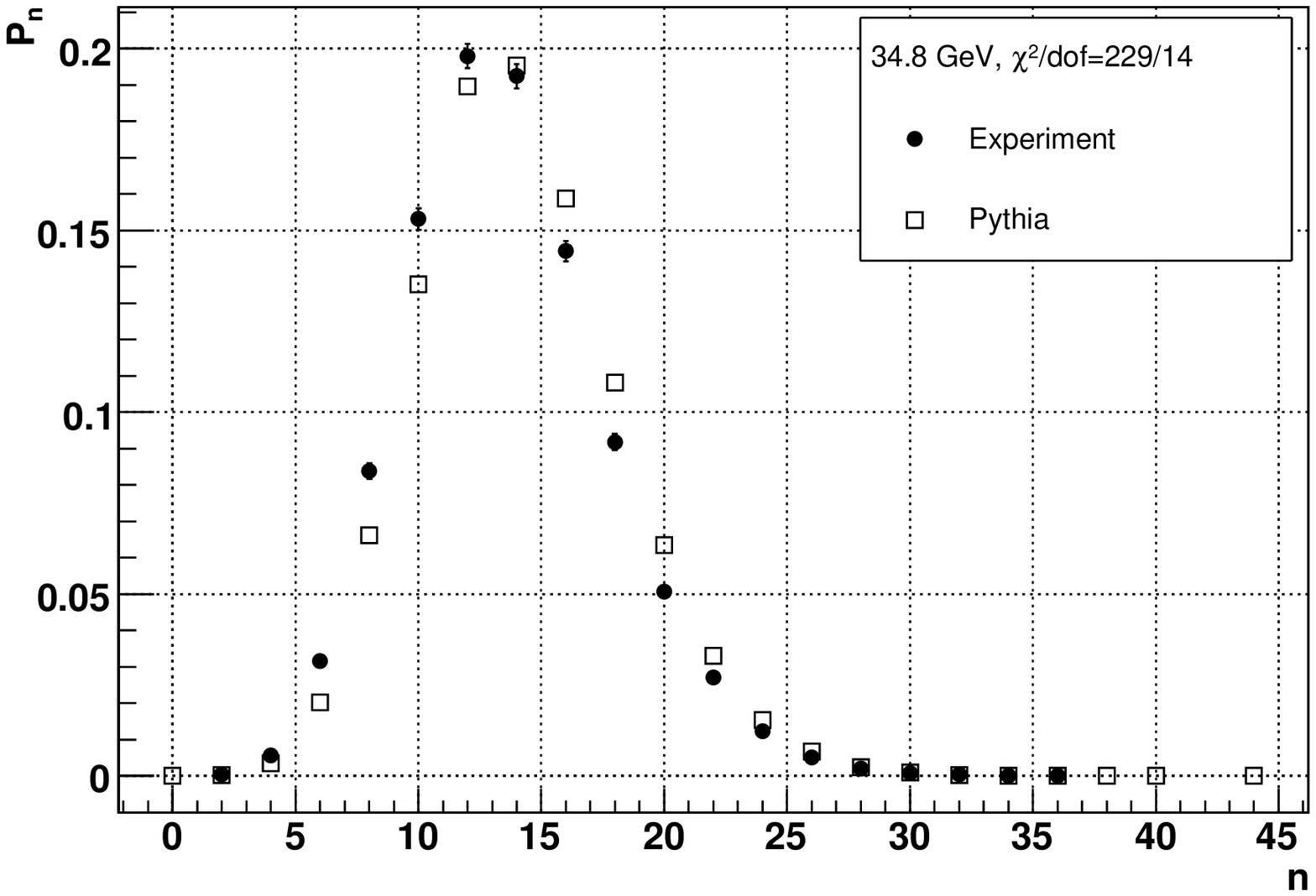}
\begin{center}
\textbf{Fig.3.}
\end{center}
\end{center}

\begin{center}
\begin{tabular}{|c|c|c|c|}
\multicolumn{4}{l}{\textbf{Table 3.} $\sqrt{s}=34.8$ GeV}\\
\hline$n$&number of events, experiment&number of events, PYTHIA&$\chi^2$ in bin\\
\hline0 -- 2&23.62&22.93&0.00\\
\hline4&302.89&302.20&0.00\\
\hline6&1673.45&1697.39&0.08\\
\hline8&4427.16&5009.39&19.14\\
\hline10&8090.16&9128.50&33.12\\
\hline12&10456.88&11249.92&14.78\\
\hline14&10167.52&10200.36&0.03\\
\hline16&7625.51&7230.66&5.23\\
\hline18&4850.98&4271.90&18.46\\
\hline20&2674.51&2160.62&27.65\\
\hline22&1434.97&971.79&43.62\\
\hline24&652.37&381.18&33.72\\
\hline26&273.30&141.01&19.60\\
\hline28&104.45&47.23&9.52\\
\hline30 -- 36&74.28&16.91&4.48\\
\hline \multicolumn{4}{|c|}{total $\chi^2$/degrees of freedom = 229/14}\\
\hline
\end{tabular}
\end{center}

\newpage
\subsection{$\sqrt{s}=91.2$ GeV}
Experimentally there were obtained 248\,100 events
(L3)~\cite{bib:5}. The experimental values of $P_n$ and the
corresponding probabilities obtained by PYTHIA are shown in
Fig.4a. Both statistical and systematical errors are included. The
$\chi^2$ is presented in Table 4a.

Unlike the previous ones, the ratio of $\chi^2$/degrees of freedom
= 1155/25 is very large for the results of L3 col., although
visually the experimental and generated points are in good
agreement (especially if one takes the curve fitting the generated
points but not the points themselves, as it was done in some
articles, for example~\cite{bib:5}, \cite{bib:6}). This is bound
to very large statistics in the L3 col.  and its small errors. The
multiplicity distributions for the same energy obtained in
DELPHI~\cite{bib:7} (25\,364 events) and OPAL~\cite{bib:8}
(82\,941 events) experiments  are given below (Fig.4b, Table 4b
and Fig.4c, Table 4c accordingly). In these experiments statistics
is lower and the values of $\chi^2$/degrees of freedom are
correspondingly smaller. The result of multiplicity distributions
from all collaborations combined together is given in Fig.4d.

\begin{center}
\includegraphics[height=115mm,width=155mm]{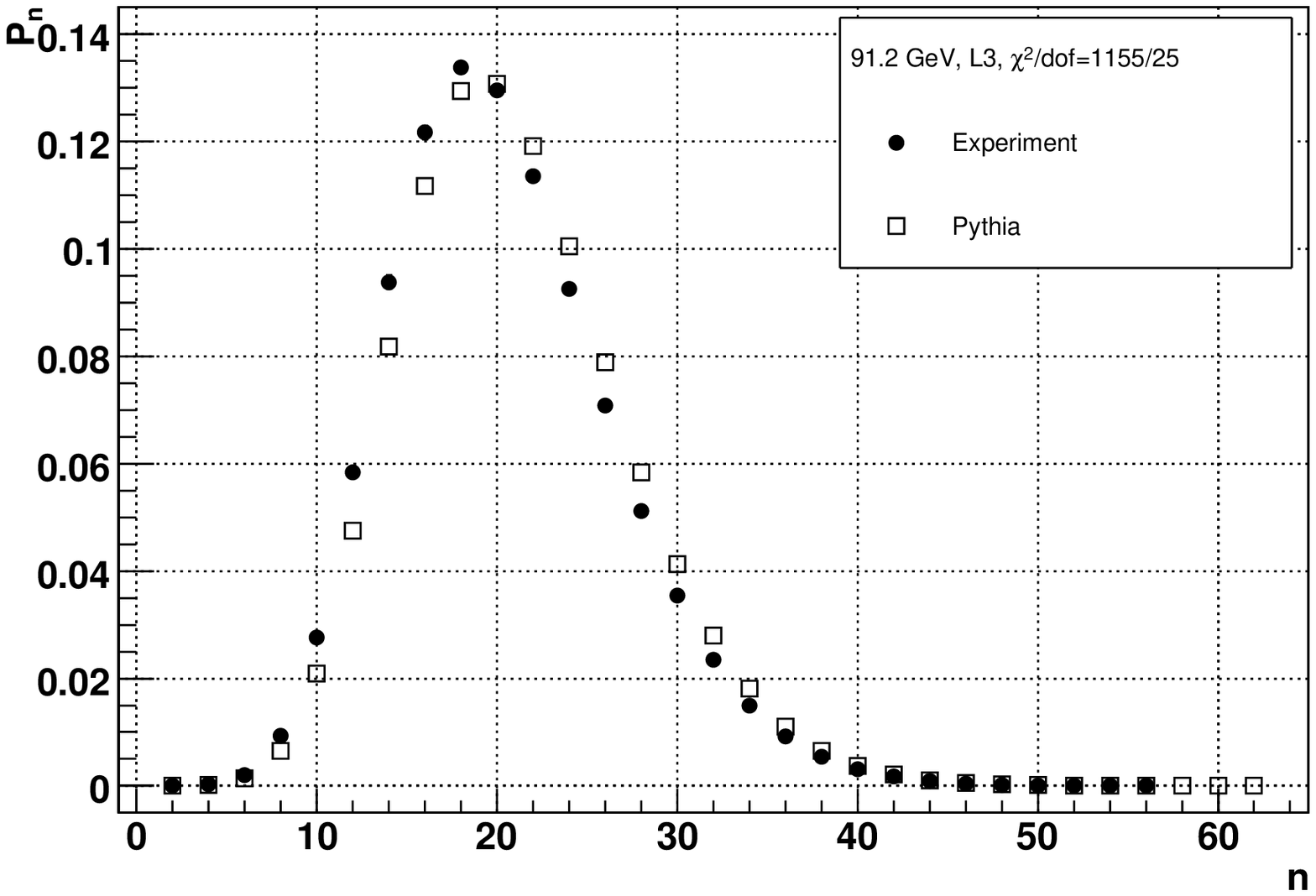}
\begin{center}
\textbf{Fig.4a.}
\end{center}
\end{center}

\begin{center}
\begin{tabular}{|c|c|c|c|}
\multicolumn{4}{l}{\textbf{Table 4a.} $\sqrt{s}=91.2$ GeV, L3 col.}\\
\hline$n$&number of events, experiment&number of events, PYTHIA&$\chi^2$ in bin\\
\hline2 -- 4&70.96&39.94&21.15\\
\hline6&509.60&343.37&54.22\\
\hline8&2314.28&1602.97&218.37\\
\hline10&6852.77&5177.60&203.97\\
\hline12&14495.49&11806.33&151.89\\
\hline14&23280.71&20305.99&90.10\\
\hline16&30198.48&27721.70&48.02\\
\hline18&33190.57&32102.16&8.85\\
\hline20&32138.63&32437.34&0.96\\
\hline22&28183.66&29575.26&24.48\\
\hline24&22970.59&24935.54&104.10\\
\hline26&17573.42&19586.50&88.30\\
\hline28&12708.92&14481.85&56.91\\
\hline30&8805.07&10245.29&28.70\\
\hline32&5840.52&6955.73&21.36\\
\hline34&3703.64&4480.69&13.01\\
\hline36&2280.04&2734.06&7.44\\
\hline38&1338.50&1619.84&5.33\\
\hline40&773.82&920.45&3.18\\
\hline42&430.45&514.31&1.97\\
\hline44&224.03&258.77&0.76\\
\hline46&112.14&134.97&0.81\\
\hline48&56.57&68.23&0.51\\
\hline50&25.31&31.51&0.35\\
\hline52&12.65&12.90&0.00\\
\hline54 -- 62&8.93&6.70&0.07\\
\hline \multicolumn{4}{|c|}{total $\chi^2$/degrees of freedom = 1155/25}\\
\hline
\end{tabular}
\end{center}

\newpage
\begin{figure}[!h]
\begin{center}
\includegraphics[height=115mm,width=155mm]{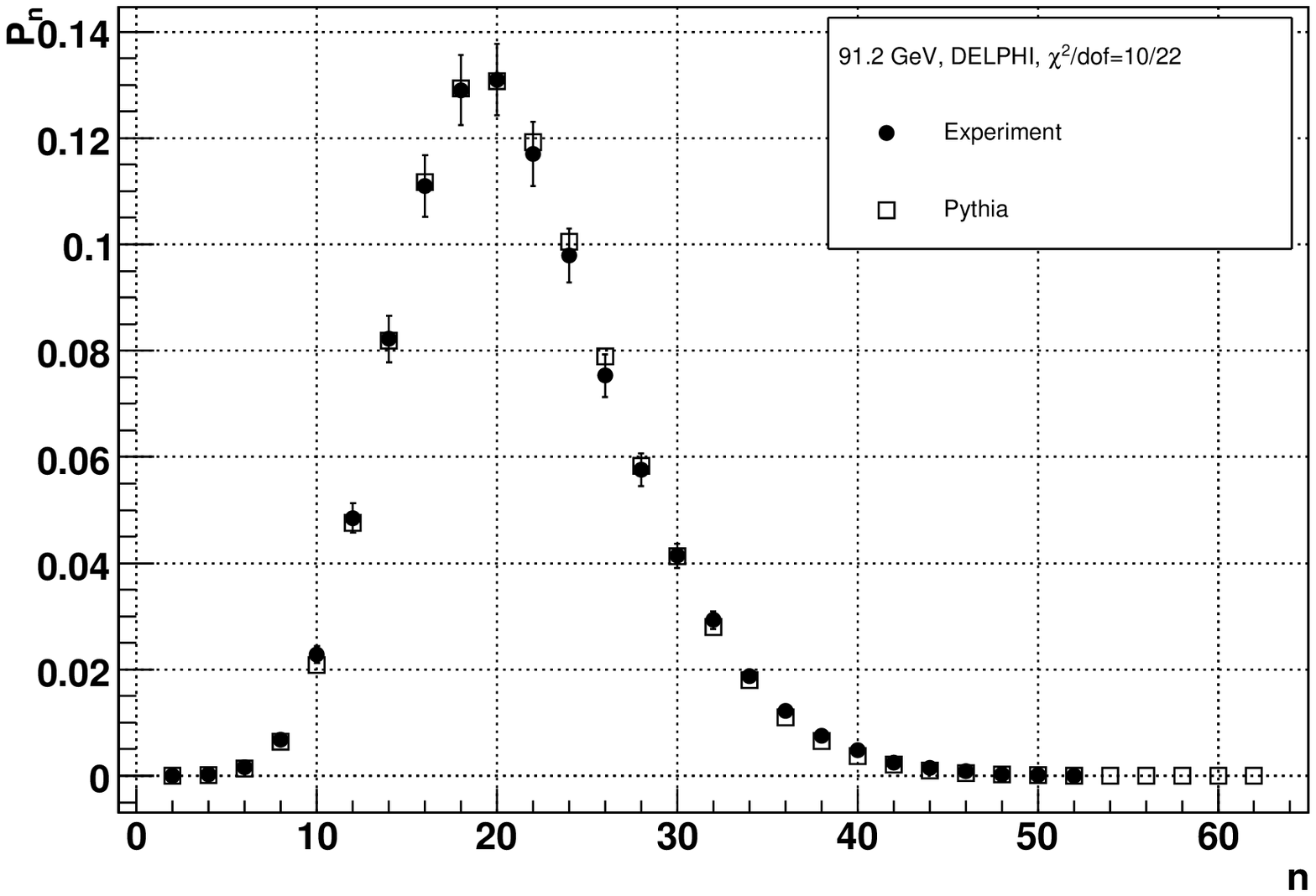}
\textbf{Fig.4b.}
\end{center}
\end{figure}

\newpage
\begin{table}[!h]
\begin{tabular}{|c|c|c|c|}
\multicolumn{4}{l}{\textbf{Table 4b.} $\sqrt{s}=91.2$ GeV, DELPHI col.}\\
\hline$n$&number of events, experiment&number of events, PYTHIA&$\chi^2$ in bin\\
\hline2 -- 6&45.91&39.19&0.23\\
\hline8&170.95&163.88&0.14\\
\hline10&578.30&529.32&1.10\\
\hline12&1230.15&1207.00&0.09\\
\hline14&2084.92&2075.94&0.01\\
\hline16&2815.40&2834.07&0.01\\
\hline18&3271.96&3281.90&0.00\\
\hline20&3322.68&3316.17&0.00\\
\hline22&2967.59&3023.57&0.12\\
\hline24&2483.14&2549.23&0.23\\
\hline26&1909.91&2002.39&0.70\\
\hline28&1460.97&1480.52&0.05\\
\hline30&1050.07&1047.41&0.00\\
\hline32&743.17&711.11&0.40\\
\hline34&476.84&458.07&0.28\\
\hline36&309.44&279.51&1.30\\
\hline38&191.50&165.60&1.83\\
\hline40&121.24&94.10&1.00\\
\hline42&63.66&52.58&0.43\\
\hline44&36.27&26.45&0.92\\
\hline46&20.80&13.80&1.16\\
\hline48&5.07&6.98&0.39\\
\hline50 -- 62&4.31&5.22&0.02\\
\hline \multicolumn{4}{|c|}{total $\chi^2$/degrees of freedom = 10/22}\\
\hline
\end{tabular}
\end{table}

\newpage
\begin{figure}[!h]
\begin{center}
\includegraphics[height=115mm,width=155mm]{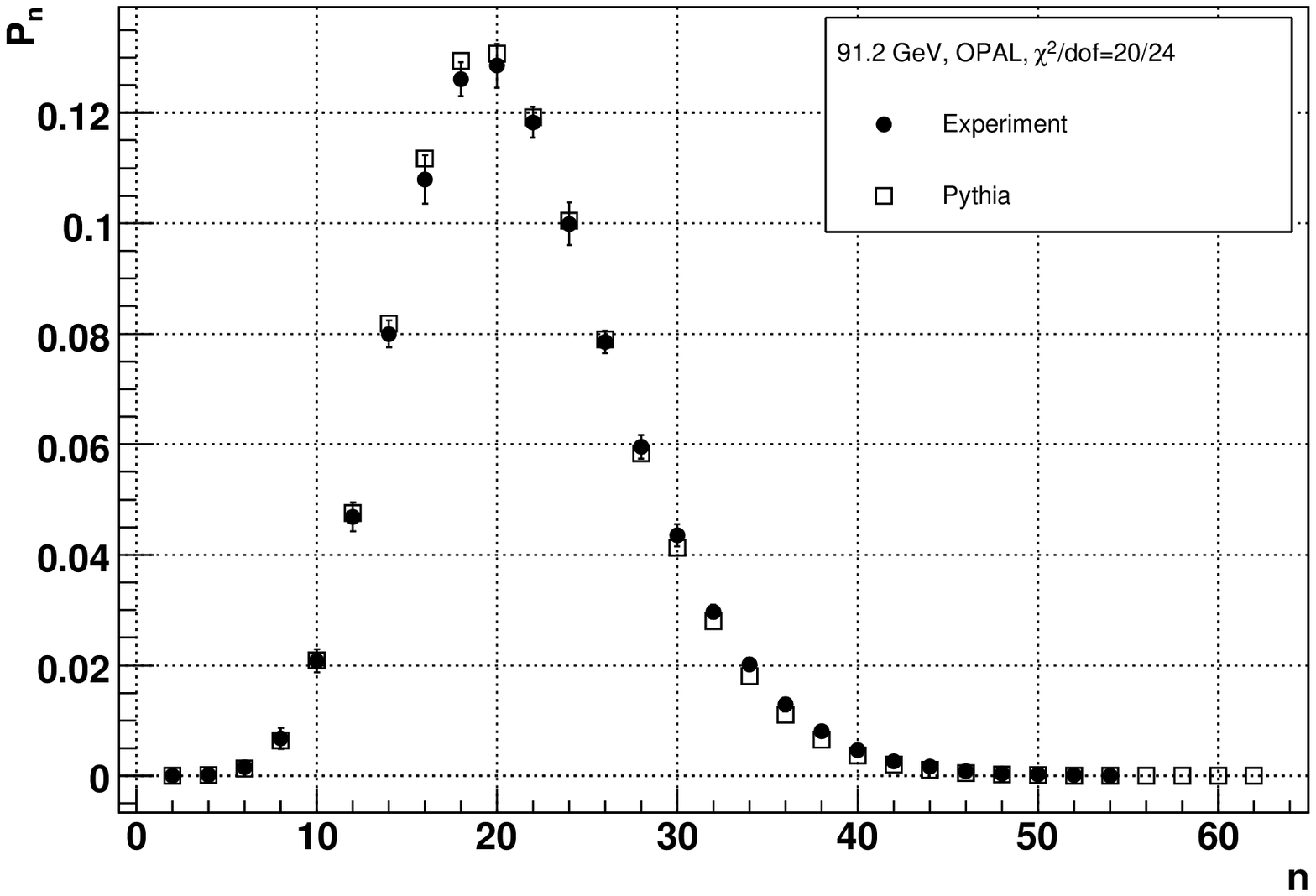}
\textbf{Fig.4c.}
\end{center}
\end{figure}

\newpage
\begin{table}[!h]
\begin{tabular}{|c|c|c|c|}
\multicolumn{4}{l}{\textbf{Table 4c.} $\sqrt{s}=91.2$ GeV, OPAL col.}\\
\hline$n$&number of events, experiment&number of events, PYTHIA&$\chi^2$ in bin\\
\hline2 -- 4&14.10&13.35&0.00\\
\hline6&132.71&114.79&0.04\\
\hline8&564.00&535.88&0.03\\
\hline10&1725.17&1730.90&0.00\\
\hline12&3889.93&3946.91&0.06\\
\hline14&6635.28&6788.39&0.48\\
\hline16&8949.33&9267.50&0.71\\
\hline18&10458.86&10731.90&0.99\\
\hline20&10657.92&10843.96&0.29\\
\hline22&9811.92&9887.15&0.09\\
\hline24&8285.81&8336.07&0.02\\
\hline26&6510.87&6547.86&0.04\\
\hline28&4934.99&4841.35&0.24\\
\hline30&3607.93&3425.05&1.06\\
\hline32&2463.35&2325.33&1.41\\
\hline34&1675.41&1497.91&2.75\\
\hline36&1069.94&914.01&2.08\\
\hline38&671.82&541.52&3.58\\
\hline40&389.82&307.71&2.16\\
\hline42&215.65&171.94&1.01\\
\hline44&141.00&86.51&1.43\\
\hline46&73.82&45.12&0.68\\
\hline48&34.84&22.81&0.38\\
\hline50&20.74&10.53&0.47\\
\hline52 -- 62&12.44&6.55&0.20\\
\hline \multicolumn{4}{|c|}{total $\chi^2$/degrees of freedom = 20/24}\\
\hline
\end{tabular}
\end{table}

\newpage
\begin{figure}[!h]
\includegraphics[height=115mm,width=155mm]{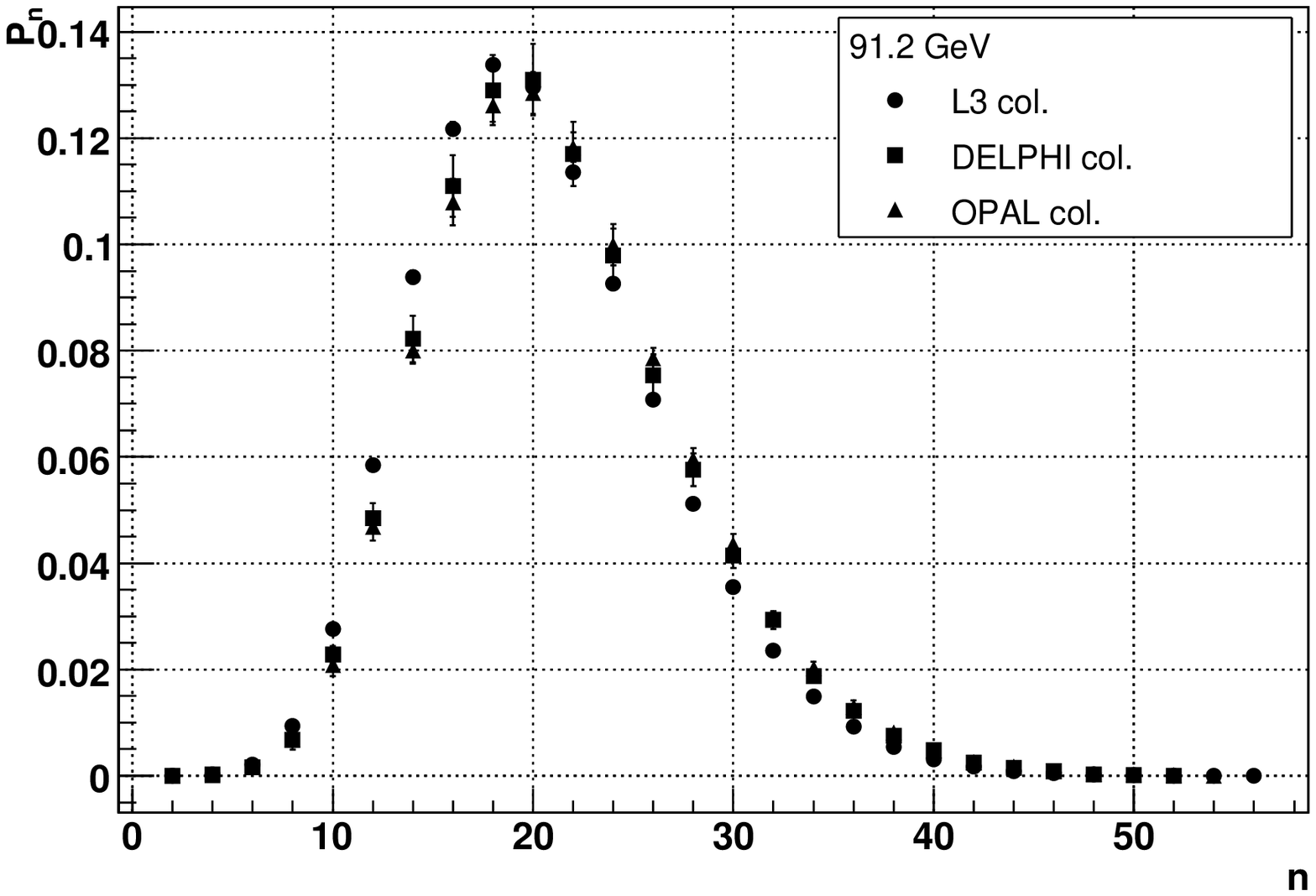}
\textbf{Fig.4d.} The comparison between the results of different
experiments at the energy of $Z^0$ peak
\end{figure}

\newpage
\subsection{$\sqrt{s}=188.6$ GeV}
Experimentally there were obtained 4\,479 events
(L3)~\cite{bib:5}. The experimental values of $P_n$ and the
corresponding probabilities obtained by PYTHIA are shown in Fig.5.
Both statistical and systematical errors are included. The
$\chi^2$ is presented in Table 5.

\begin{figure}[!h]
\begin{center}
\includegraphics[height=115mm,width=155mm]{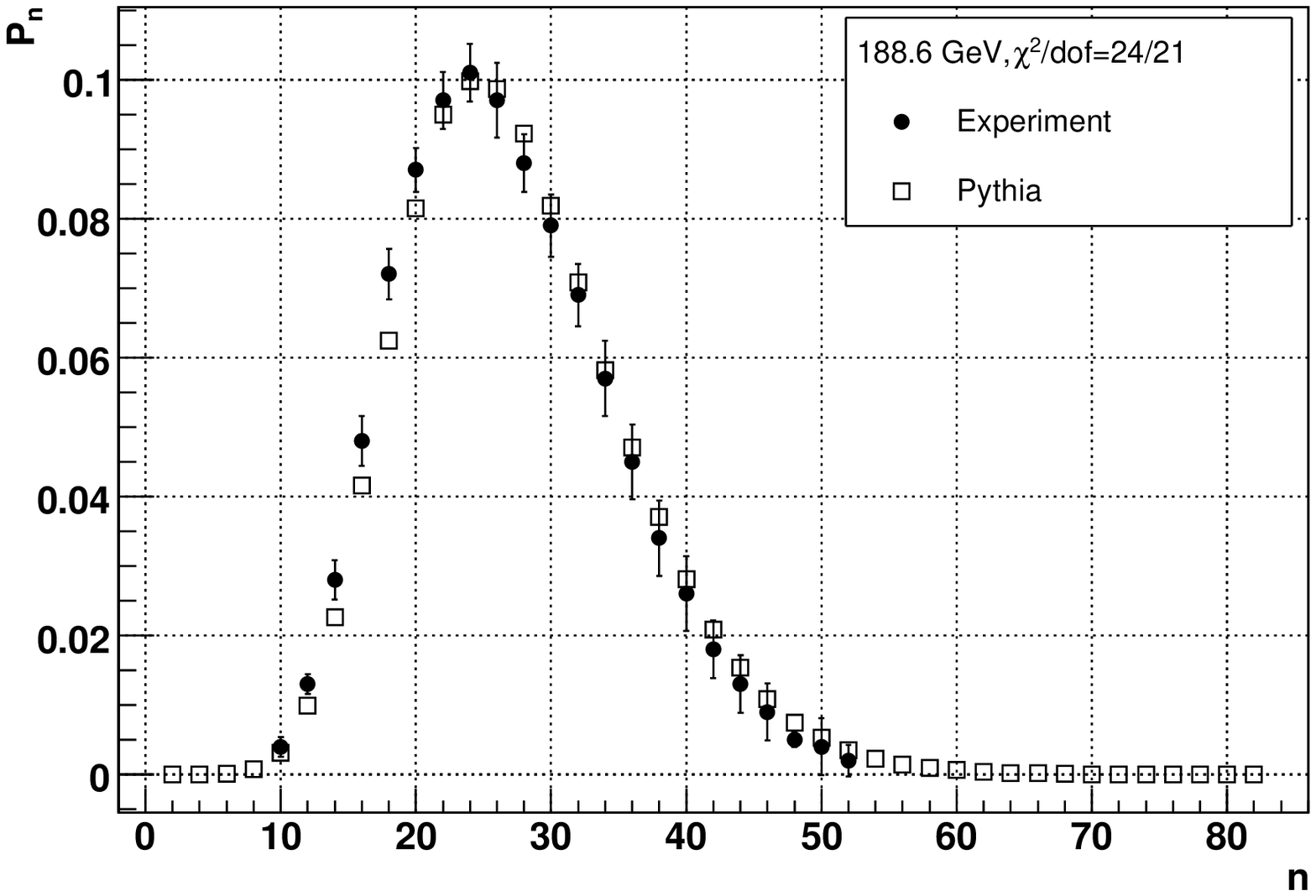}
\textbf{Fig.5.}
\end{center}
\end{figure}

\newpage
\begin{table}[!h]
\begin{tabular}{|c|c|c|c|}
\multicolumn{4}{l}{\textbf{Table 5.} $\sqrt{s}=188.6$ GeV}\\
\hline$n$&number of events, experiment&number of events, PYTHIA&$\chi^2$ in bin\\
\hline2 -- 10&17.92&18.16&0.00\\
\hline12&58.23&44.23&2.32\\
\hline14&125.41&101.23&2.23\\
\hline16&214.99&186.12&1.87\\
\hline18&322.49&279.53&3.41\\
\hline20&389.67&364.89&1.09\\
\hline22&434.46&425.13&0.11\\
\hline24&452.38&447.01&0.04\\
\hline26&434.46&441.80&0.05\\
\hline28&394.15&413.00&0.47\\
\hline30&353.84&366.51&0.21\\
\hline32&309.05&317.03&0.09\\
\hline34&255.30&260.58&0.03\\
\hline36&201.56&210.87&0.11\\
\hline38&152.29&165.96&0.25\\
\hline40&116.45&125.96&0.13\\
\hline42&80.62&93.32&0.37\\
\hline44&58.23&68.70&0.27\\
\hline46&40.31&48.67&0.18\\
\hline48&22.40&33.54&2.32\\
\hline50&17.92&23.68&0.09\\
\hline52 -- 82&8.96&43.07&8.12
\\
\hline \multicolumn{4}{|c|}{total $\chi^2$/degrees of freedom = 24/21}\\
\hline
\end{tabular}
\end{table}

\newpage
\subsection{$\sqrt{s}=206.2$ GeV}
Experimentally there were obtained 4\,146 events
(L3)~\cite{bib:5}. The experimental values of $P_n$ and the
corresponding probabilities obtained by PYTHIA are shown in Fig.6.
Both statistical and systematical errors are included. The
$\chi^2$ is presented in Table 6.

\begin{figure}[!h]
\begin{center}
\includegraphics[height=115mm,width=155mm]{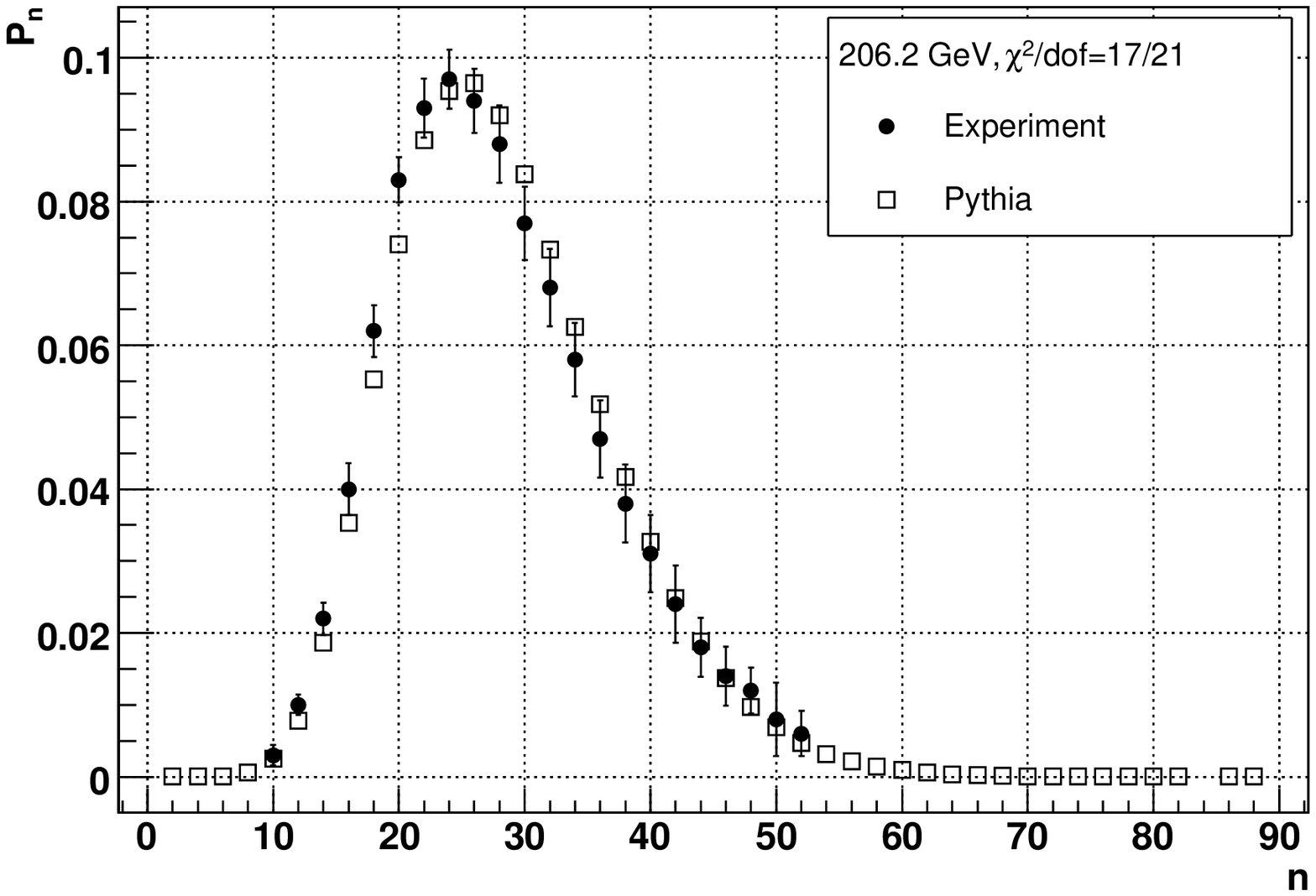}
\textbf{Fig.6.}
\end{center}
\end{figure}

\newpage
\begin{table}[!h]
\begin{tabular}{|c|c|c|c|}
\multicolumn{4}{l}{\textbf{Table 6.} $\sqrt{s}=206.2$ GeV}\\
\hline$n$&number of events, experiment&number of events, PYTHIA&$\chi^2$ in bin\\
\hline2 -- 10&12.44&13.43&0.02\\
\hline12&41.46&32.33&1.25\\
\hline14&91.21&77.09&1.22\\
\hline16&165.84&146.40&1.02\\
\hline18&257.05&229.01&1.74\\
\hline20&344.12&306.95&2.89\\
\hline22&385.58&366.97&0.53\\
\hline24&402.16&395.36&0.07\\
\hline26&389.72&400.01&0.14\\
\hline28&364.85&381.23&0.31\\
\hline30&319.24&347.40&1.00\\
\hline32&281.93&303.93&0.60\\
\hline34&240.47&259.44&0.51\\
\hline36&194.86&214.66&0.55\\
\hline38&157.55&172.92&0.35\\
\hline40&128.53&135.30&0.07\\
\hline42&99.50&103.13&0.02\\
\hline44&74.63&78.00&0.03\\
\hline46&58.04&56.78&0.00\\
\hline48&49.75&40.26&0.42\\
\hline50&33.17&28.50&0.05\\
\hline52 -- 88&24.88&56.91&4.48
\\
\hline \multicolumn{4}{|c|}{total $\chi^2$/degrees of freedom = 24/21}\\
\hline
\end{tabular}
\end{table}

\newpage
\section{Conclusion}
As it was shown, PYTHIA  describes well the experimental data on
the multiplicity distribution in $e^+e^-$ annihilation to hadrons.
So it can be used as a reliable method of theoretical studies.

As it was noticed in PYTHIA manual~\cite{bib:1}, the default
values of all its parameters were tuned to describe the events at
the energy of $Z^0$ peak. The ratio of $\chi^2$/degrees of freedom
is equal to 10/22 for DELPHI col. with 25\,364 events, 20/24 for
OPAL col. with 82\,941 events and it is surprisingly large,
1155/25, for L3 col. with 248\,100 events. The ratio is
approximately equal to 1 for other discussed energies, although
for 34.8 it is 229/14 (52\,832 events, TASSO col.). This behavior
of $\chi^2$ is quite obvious. Thus, more subtle tuning is
necessary for the values of PYTHIA parameters, so that PYTHIA will
be able to predict quantity results.

Author is grateful for V.\,A.~Abramovsky for useful discussions.

This work was partially supported by RFBR grant 07-07-96410-r
center-a.

\end{document}